\documentclass[aps,prl,twocolumn,groupedaddress,showpacs]{revtex4}

\usepackage{graphicx}% Include figure files
\usepackage{dcolumn}% Align table columns on decimal point
\usepackage{bm}% bold math

\begin{document}

%\preprint{APS/123-QED}

\title{Conductance Anomaly and Fano Factor Reduction in Quantum Point
Contacts}

\author{Shuji Nakamura*, Masayuki Hashisaka*, Yoshiaki Yamauchi, Shinya
Kasai, Teruo Ono, and Kensuke Kobayashi} \affiliation{Institute for
Chemical Research, Kyoto University, Uij,Kyoto 611-0011, Japan.}
\affiliation{*These authors contributed equally to this work.}


\date{\today}% It is always \today, today,

\begin{abstract}
We report an experimental study on the shot noise as well as the dc
transport properties of a quantum point contact (QPC) whose conductance
anomaly can be tuned electrostatically by the gate electrodes.  By
controlling the single QPC so that it has no anomaly or an anomaly at
$0.5 G_0$, $0.8 G_0$ or $0.9 G_0$ ($G_0 = 2e^2/h$), we prove that the
anomaly always accompanies the Fano factor reduction due to the
asymmetric transmission of the two spin-dependent channels for the
conductance lower than $G_0$.  For the QPC tuned to have the anomaly
at $0.5 G_0$ the channel asymmetry is found to be as large as 67 \% with
the spin gap energy gradually evolving as the conductance increases.
\end{abstract}



\maketitle The conductance of electron ($G$) through a quantum point
contact (QPC) is a beautiful manifestation of the Landauer
formula~\cite{WeesPRL1988,WharamJPC1988}.  Just before the pinch-off
only the transmission of the last conducting mode ($T_0$) contributes to
the conductance resulting in $G=G_0 T_0$, where $G_0 = 2e^2/h \sim (12.9
\mbox{~k}\Omega)^{-1}$. In spite of a naive expectation that $T_0$
monotonously behaves as a function of the opening of QPC, there often
appears a shoulder structure (``anomaly'') at $G \sim 0.7
G_0$~\cite{ThomasPRL1996}. A number of conductance measurements have
suggested that this ``0.7 anomaly'', which may appear between at
$0.5G_0$ and at $G_0$, is caused by the lifting of the spin degree of
freedom even in a zero magnetic
field~\cite{ThomasPRL1996,ThomasPRB1998,KristensenPRB2000,CronenwettPRL2002,
RochePRL2004,DiCarloPRL2006}, while its microscopic origin remains to be
clarified in spite of intensive theoretical
efforts~\cite{WangPRB1996,BruusPE2001,MeirPRL2002,MatveevPRL2004,
ReillyPRB2005,RejecNature2006,GolubPRL2006,LasslPRB2007,SzaferPRL1989}.
When the spin is explicitly taken into account, the Landauer formula for
the last conducting mode yields $G = G_0/2 \sum_\sigma \tau_\sigma$,
where $\tau_\sigma$ is the transmission of the last channel with spin
$\sigma$ ($= \uparrow \mbox{or} \downarrow$).  The above argument means
that the single particle picture that $T_0 = \tau_\uparrow =
\tau_\downarrow$ does not hold true in the anomaly case so that two
spin-dependent channels asymmetrically contribute to the conductance to
cause the anomaly. The conductance measurement alone, however, is not
sufficient to address this issue as it only gives the averaged
transmission.

The quantum shot noise is a powerful probe to provide more detailed
information on the transmission as it results from the partition process
of electrons~\cite{BlanterPR2000}. When the DC current $I$ is fed to QPC
the shot noise yields the current fluctuation $S_I = 2eIF$ in the
zero-temperature zero-frequency limit, where $F$ is the Fano factor to
characterize the above partition process. The mesoscopic scattering
theory~\cite{BlanterPR2000} predicts $F = {\sum_\sigma \tau_\sigma
(1-\tau_\sigma)}/{\sum_\sigma \tau_\sigma}$ for $G \leq G_0$ and
therefore $F$ simply equals $1 - G/G_0$ when the spin is
degenerated. While the pioneering
works~\cite{ReznikovPRL1995,KumarPRL1996,LiuNature1998} including recent
ones~\cite{GershonPRL2008,HashisakaPRB2008} overall confirmed the
theory~\cite{LesovikJETP1989,ButtikerPRL1990,MartinPRB1992}, some recent
measurements~\cite{RochePRL2004,DiCarloPRL2006} reported that $F$ is
smaller than this value at the 0.7 anomaly. This is clear evidence of
the asymmetric contribution of the two spin-dependent channels to the
conductance and the shot noise, because the inequality $F = \left(
(1-\tau_\uparrow)\tau_\uparrow+(1-\tau_\downarrow) \tau_\downarrow
\right) /(\tau_\uparrow +\tau_\downarrow) < 1-G/G_0$ always holds for
$\tau_\uparrow \neq \tau_\downarrow$~\cite{RochePRL2004,DiCarloPRL2006}.
In other words, the channel asymmetry $A \equiv |\tau_\uparrow
-\tau_\downarrow|/ (\tau_\uparrow +\tau_\downarrow)$ is finite.
Although here the electron correlation is only included in the
Hartree-Fock level in the framework of the Landauer-B{\"u}ttiker
formalism, the Fano factor reduction due to the asymmetry serves as an
important signature of the correlation. While the recent first-principle
calculation successfully proves a finite asymmetry for the QPC with a
given potential shape~\cite{RejecNature2006}, there have been no
experiments to address how large it can be depending on the actual
potential shapes of QPC.

Here we report an experimental study on the relation between the Fano
factor reduction and the conductance anomaly. By using the QPC whose
anomaly can be tuned between at $0.5G_0$ and $G_0$ electrostatically, we
show that the conductance anomaly accompanies with the Fano factor
reduction and the zero-bias peak in the differential conductance. We
also show that when there is no anomaly, the Fano factor perfectly
agrees with the conventional shot noise theory. The asymmetry and the
spin gap between the spin-dependent channels of QPC when $G \leq G_0$
are discussed.

\begin{figure}[htb]
\begin{center}
\center \includegraphics[width=0.80\linewidth]{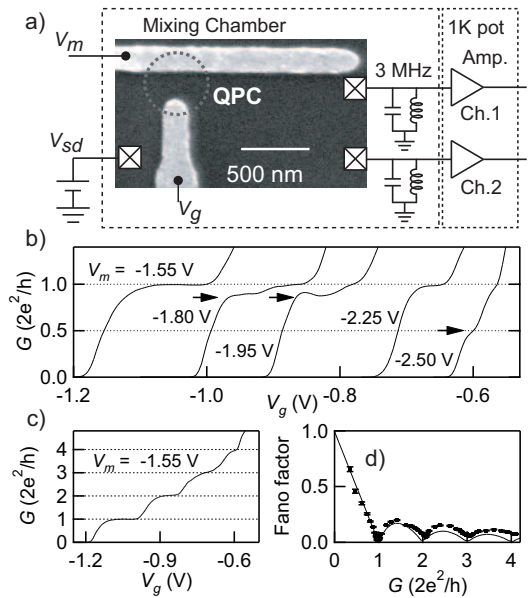} \caption{
a) Schematic diagram of the measurement setup with the scanning electron
microscope image of the sample fabricated on the GaAs/AlGaAs 2DEG.  Two
gates electrodes, namely the main gate ($V_m$) and the side gate ($V_g$)
are used to define QPC. The conductance measurement was performed by a
standard lock-in technique with an excitation voltage of 10~$\mu$V.  For
the shot noise measurement, the cross-correlation technique was used
with the two cryogenic amplifiers.  b) Typical examples of the QPC
conductance are shown as a function of $V_g$ for several $V_m$
values. For $V_m = -1.80$, -1.95, and -2.50~V, the conductance anomaly
appears as indicated by the arrows. c) When $V_m = -1.55$~V four
conductance plateaus are clearly observed without any signature of the
anomaly. d) The obtained Fano factor for $V_m = -1.55$~V presented in
points with error bars behaves just as the standard theory predicts
(shown in the solid curve).  } \label{Fig1}
\end{center}
\end{figure}

Figure 1a shows the scanning electron micrograph of our QPC fabricated
on the two-dimensional electron gas (2DEG) in the AlGaAs/GaAs
heterostructure and the experimental setup for the shot noise
measurement. The current fluctuation at 3.0 MHz defined by the resonant
(inductor-capacitor) circuit is measured through the home-made cryogenic
amplifier to obtain the shot
noise~\cite{DiCarloPRL2006,HashisakaPRB2008}. To increase the resolution
of the noise spectrum, the cross-correlation technique is
used~\cite{RochePRL2004,KumarPRL1996} with two sets of the resonant
circuit and the amplifier~\cite{DiCarloPRL2006}. The experiment was
performed in the dilution refrigerator whose base temperature is 45~mK
and the electrical temperature ($T_e$) in the equilibrium states was
calibrated to be 125~mK by measuring the thermal noise. A slight
magnetic field (0.2~T) was applied perpendicular to the 2DEG as
performed before~\cite{CronenwettPRL2002,DiCarloPRL2006,KumarPRL1996}.

The present QPC is defined by the main gate and the side gate to which
the voltages $V_m$ and $V_g$ are applied, respectively (Fig.~1a). By
modifying the two gate voltages the curvature of QPC is varied; when
$V_m$ is negatively large, the constriction potential of QPC has a large
curvature with small $|V_g|$ and vice versa. Figure 1b shows the QPC
profile as a function of $V_g$ for several $V_m$'s. In going from the
small curvature ($V_m= -1.55$~V) to the large one ($V_m= -2.50$~V), the
conductance plateau becomes shorter as expected from the saddle point
model for QPC~\cite{ButtikerPRB1990}. While there is no anomaly for
$V_m= -1.55$ and -2.25~V, an anomaly appears at $V_m= -1.80$~V and -1.95
V around $G = 0.9G_0$ and at $V_m = -2.50$ V around $G = 0.5G_0$.

The current fluctuation $S_I$ at the source-drain voltage ($V_{sd}$) was
analyzed based on the following formula~\cite{BlanterPR2000}
\begin{equation}
S_I (V_{sd}) = 4k_BT_eG + 2FG (eV_{sd} 
\coth(\frac{eV_{sd}}{2k_BT_e}) -2k_BT_e),
\end{equation}
where $k_B$ is the Boltzmann constant.  For example, the obtained Fano
factor as a function of the conductance for $V_m= -1.55$~V, where the
four conductance plateaus are clearly observed (Fig.~1c), is plotted in
Fig. 1d. The result satisfactorily agrees with the standard theory of
the quantum shot noise expected for a simple
QPC~\cite{BlanterPR2000,ButtikerPRL1990,CommentOnHeating}.

\begin{figure*}[htb]
\begin{center}
\center
\includegraphics[width=0.80\linewidth]{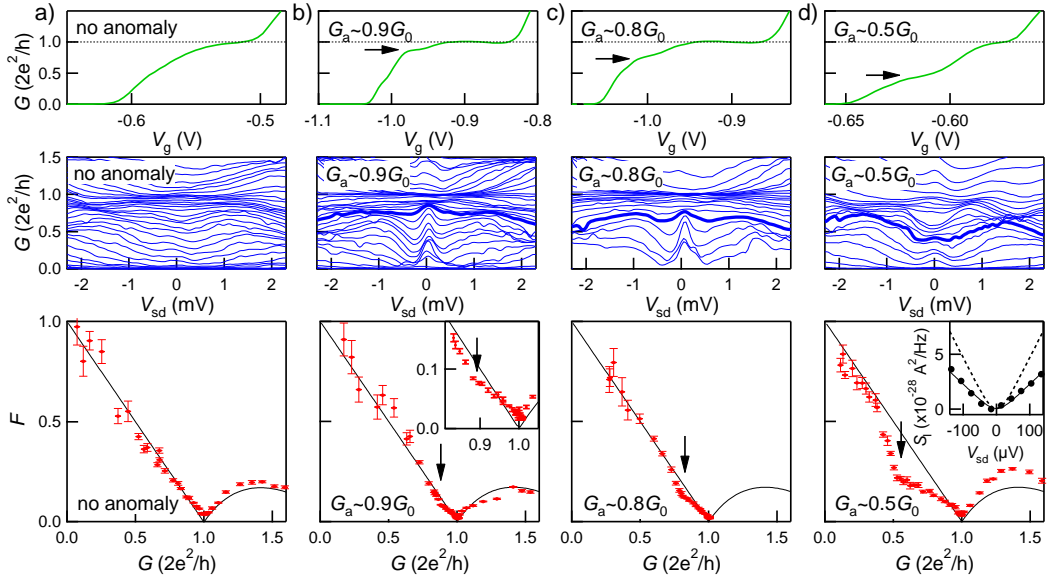}
\caption{(color online) a) The conductance ($G$), the differential
conductance, and the Fano factor ($F$) are plotted as a function of
$V_g$, $V_{sd}$, and $G$ in the upper, middle, and lower panels,
respectively. These are obtained when there is no anomaly in the QPC
profile. Each curve in the differential conductance plot is obtained for
several different $V_g$'s. We set the bias range $e|V_{sd}|\leq 150
\mu$eV for obtaining the Fano factor to minimize the effect of the
non-linearity of the conductance, while we have confirmed that a slight
variation of the bias window (100 - 250 $\mu$eV) does not affect the
result. b), c), and d) the corresponding plots for the cases when the
conductance has an anomaly around $0.9G_0$, $0.8G_0$, and $0.5G_0$. In
the inset of the bottom panel in b), the close-up of the Fano factor
behavior around the anomaly region is shown.  In the inset of the bottom
panel in d), the shot noise observed at $G = 0.5G_0$ is shown as a
function of $V_{sd}$. The solid circles represent the experimental data
with the fitted curve by Eqn (1) in a solid curve with the dashed curve
for the predicted shot noise for the non-interacting case.  }
\label{Fig2}
\end{center}
\end{figure*}

From now on we focus ourselves on the four kinds of conductance profiles
obtained for the present QPC by varying $V_m$; the QPC with ``no
anomaly'', the ones with the anomaly at $G_a \sim 0.9G_0$, at $G_a \sim
0.8G_0$ and at $G_a \sim 0.5G_0$ as shown in solid curves in the top
panels of Figs.~2a, 2b, 2c, and 2d, respectively. In the middle panels
are shown the corresponding differential conductance plots obtained at
finite $V_{sd}$ for different $V_g$'s. The Fano factor as a function of
$G$ is plotted in points for the four cases in the bottom panels of
Figs.~2a, 2b, 2c, and 2d. In all four cases, not only the conductance
quantization is observed but also the Fano factor is very close to zero
at $G = G_0$, which ensures that only the last conducting mode
contributes to the conductance when $G \leq G_0$.

For the ``no anomaly'' case shown in Fig. 2a, the conductance profile is
monotonous as a function of $V_g$ and the differential conductance has
no particular structure around $V_{sd} =0$~mV. The Fano factor increases
according to $1 - G/G_0$ (shown in a solid line) as the conductance
decreases from $G_0$; the standard shot noise theory is satisfactorily
valid~\cite{BlanterPR2000,ButtikerPRL1990}. On the other hand, for the
$G_a \sim 0.9G_0$ case (Fig.~2b), the peak structure around $V_{sd} =
0$~mV appears in the non-linear differential conductance as the
conductance decreases from $G_0$ across the anomaly at $G_a$. This
observation is consistent with the well-known zero-bias peak as often
reported for the QPC
anomaly~\cite{KristensenPRB2000,CronenwettPRL2002,DiCarloPRL2006}.
While the Fano factor obeys the theoretical curve $F = 1 - G/G_0$ (shown
in a solid curve) for $G \leq 0.8G_0$, it has a clear dip structure
around $G_a$ as seen in the inset of the bottom panel of Fig.~2b. The
reduction of the Fano factor at the anomaly is consistent with the
previous reports~\cite{RochePRL2004,DiCarloPRL2006}.

Importantly, a similar observation holds true for the $G_a \sim 0.8G_0$
and $G_a \sim 0.5G_0$ cases; the reduction of the Fano factor is most
prominent around $G = G_a$ and the zero-bias peak in the differential
conductance emerges there.  Thus, the conductance anomaly, the zero-bias
peak in the differential conductance and the reduction of the Fano
factor are all synchronous.  In the inset of the bottom panel of Fig.~2d
shown is the shot noise (excess noise on top of the thermal noise) at $G
= 0.5G_0$ for the $G_a \sim 0.5G_0$ case as a function of $V_{sd}$.  The
solid circles represent the experimental data with the fitted curve by
Eqn (1). The shot noise is indeed remarkably reduced with $F = 0.22$
from the theoretical value ($F = 0.5$).

By combining the results of the conductance $G = G_0 (\tau_\uparrow
+\tau_\downarrow)/2$ and the Fano factor $F = \left(
(1-\tau_\uparrow)\tau_\uparrow+(1-\tau_\downarrow) \tau_\downarrow
\right) /(\tau_\uparrow +\tau_\downarrow)$, we decompose the
transmission to the two channels on the assumption that the conductance
below $G_0$ is composed of two spin-dependent channels whose
transmission are given by $\tau_\uparrow$ and
$\tau_\downarrow$~\cite{ReillyPRB2005,DiCarloPRL2006}. In the bottom
panel of Fig.~3 are shown the decomposed transmissions obtained from the
data shown in Fig.~2d.  The channel asymmetry $A = |\tau_\uparrow
-\tau_\downarrow|/ (\tau_\uparrow +\tau_\downarrow)$ is also shown in
the middle panel. At the conductance plateau $G_0$, the transmissions of
both channels are close to unity and hence $A = 0$ within the
experimental accuracy. As the conductance decreases, the finite
asymmetry occurs and takes the maximum value of 67 $\pm$ 3 \% around the
anomaly $0.5G_0$. As the conductance decreases, $A$ rapidly decreases to
zero again, although $A$ cannot be sufficiently precisely determined for
$G < 0.3G_0$ due to the low resolution of the Fano factor at the lower
conductance region. In spite of this limitation, the value as large as
67 \% has enough accuracy.  For the cases of $G_a \sim 0.9G_0$ (Fig.~2b)
and $G_a \sim 0.8G_0$ (Fig.~2c), $A$ takes its maximum around the
conductance anomaly to be 20 $\pm$ 2 \% and 26 $\pm$ 3 \%, respectively.

\begin{figure}[htb]
\begin{center}
\center \includegraphics[width=0.8\linewidth]{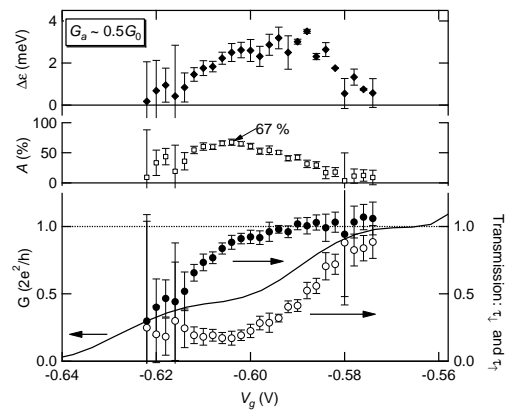}
\caption{ In the lower panel the decomposed spin-dependent transmissions
of the last channel ($\tau_\uparrow$ and $\tau_\downarrow$) for the $G_a
\sim 0.5G_0$ case is shown as a function of the side gate voltage $V_g$
with the conductance profile superposed. In the middle and the top
panels, the obtained asymmetry $A$ and the spin gap energy
$\Delta\epsilon$ are shown, respectively.}  \label{Fig3}
\end{center}
\end{figure}

The gap energy between the two channels is estimated by assuming the
energy-dependent transmission $\tau_\sigma (\epsilon) = 1/(1+e^{2\pi
(\epsilon_\sigma-\epsilon)/\hbar \omega_x})$~\cite{DiCarloPRL2006},
which is appropriate for the saddle point potential with the curvature
parallel to the current defined by
$\omega_x$~\cite{ButtikerPRB1990,SaddlePointNote}.  $\omega_x$ is
obtained by fitting the conductance profile for $G \leq 0.3 G_0$ and the
energy gap $\Delta \epsilon \equiv |\epsilon_\uparrow -
\epsilon_\downarrow|$ is calculated from $\tau_\uparrow$ and
$\tau_\downarrow$ (Fig.~3), where the lever arm extracted from the
transconductance data (in the middle panel of Fig.~2d) is used to
convert $V_g$ to the energy.  The obtained $\Delta \epsilon$ is plotted
in the top panel of Fig.~3. In the $G_a \sim 0.5 G_0$ case
($\hbar\omega_x = 4.4$ meV), $\Delta \epsilon$ almost linearly increases
as a function of $V_g$ from 0~meV (at $G = 0.3G_0$) to 3.5~meV (at
$0.8G_0$), being in agreement with the phenomenological model with the
energy gap depending on the gate or the electron
density~\cite{ReillyPRB2005}.  At $G = 0.5G_0$, $\Delta \epsilon = 2.5$
meV much larger than $k_B T_e$, which agrees with our observation that
the conductance profile is only weakly dependent on the temperature
below 1~K and the magnetic field below 3~T when the conductance profile
has an anomaly around $0.5 G_0$.  In the same way $\Delta \epsilon =
2.2$~meV at $G = G_a$ for the $G_a \sim 0.8$ case ($\hbar\omega_x = 3.5$
meV), and $\Delta \epsilon = 1.8$~meV at $G = G_a$ for the $G_a \sim
0.9$ case ($\hbar\omega_x = 5.2$ meV).  For comparison, $\Delta \epsilon
= 0.5$~meV at $G =0.7 G_0$ for the previous
experiment~\cite{DiCarloPRL2006}.

The obtained values of the asymmetry $A$ is remarkably large especially
for the $G_a \sim 0.5$ case since the similar analysis gives $A = 14$~\%
at the conductance anomaly for the data reported
before~\cite{DiCarloPRL2006}. Some theories discuss that the QPC anomaly
is attributed to the spontaneous spin polarization due to the exchange
interaction in QPC~\cite{WangPRB1996,LasslPRB2007}. If
it would be the case, as is the case in QPC in the two-dimensional
hole gas~\cite{RokhinsonPRL2006}, the asymmetry $A$ would give the spin
polarization of the electrons at QPC. In this case, 67 \% is
comparable to that of the typical ferromagnets like permalloy.  On the
other hand, other theoretical
approaches~\cite{BruusPE2001,MeirPRL2002,MatveevPRL2004,GolubPRL2006} do
not require a finite magnetic moment at QPC; for example, the Kondo
mechanism successfully reproduces the shot noise behavior for the 0.7
anomaly case~\cite{GolubPRL2006}.  To determine which the case is, the
frequency dependence of the shot noise in the higher frequency range is
necessary.

To conclude, we show that the anomaly accompanies the Fano factor
reduction and the zero-bias peak in the differential conductance. After
proving that the Fano factor is close to zero at $G=G_0$, we discuss the
asymmetric transmission of the two spin-dependent channels for $G \leq
G_0$.  The asymmetry for the $0.5G_0$ anomaly case is as large as 67 \%
with the spin gap energy gradually evolving from 0 to 3.5~meV as the
conductance increases.  Further clarification on the relation between
the asymmetry and the QPC potential will open up a new strategy to tune
the spin-dependent transport in low-dimensional systems.

We appreciate fruitful comments from Y. Avishai, M. B{\"u}ttiker,
Y. C. Chung, C. J. B. Ford, L. Glazman, R. Leturcq, T. Martin, Y. Meir,
P. Roche, C. Texier, Y. Tokura, and M. Ueda. This work is supported by
KAKENHI, Yamada Science Foundation, and Matsuo Science Foundation.


\begin{thebibliography}{99}

\bibitem{WeesPRL1988} B.~J.~van~ Wees, \textit{et al.}
Phys. Rev. Lett. {\bf 60,} 848 (1988). 

\bibitem{WharamJPC1988} D.~A.~Wharam {\it et al.}, J. Phys. C {\bf 21,} 
L209 (1988). 

\bibitem{ThomasPRL1996} K.~J.~Thomas {\it et al.,} Phys. Rev. Lett. {\bf
77,} 135 (1996).

\bibitem{ThomasPRB1998} K.~J.~Thomas {\it et al.,} Phys. Rev. B {\bf
58,} 4846 (1998); K.~J.~Thomas, {\it et al.,} Phys. Rev. B {\bf 61,}
R13365 (2000).; S.~Nuttinck {\it et al.,} Jpn. J. Appl. Phys.  {\bf 39,} L655
(2000); D.~J.~Reilly {\it et al.,} Phys. Rev. Lett. {\bf 89,} 246801
(2002); R.~Crook {\it et al.,} Science {\bf 312,} 1359 (2006);
Y.~Yoon {\it et al.,} Phys. Rev. Lett. {\bf 99,} 136805 (2007); Y.~Chung
{\it et al.,} Phys. Rev. B {\bf 76,} 035316 (2007).
 
\bibitem{KristensenPRB2000} A.~Kristensen {\it et al.,} Phys. Rev. B {\bf 62,} 10950
(2000).

\bibitem{CronenwettPRL2002} S.~M.~Cronenwett {\it et al.,} Phys. Rev.  Lett. {\bf 88,}
226805 (2002).

\bibitem{RochePRL2004} P.~Roche {\it et al.,} Phys. Rev. Lett. {\bf 93,} 
116602 (2004).

\bibitem{DiCarloPRL2006} L.~DiCarlo {\it et al.,} Phys. Rev. Lett. 
{\bf 97,} 036810 (2006). 

\bibitem{WangPRB1996} C.~K.~Wang and K.~F.~Berggren, Phys. Rev. B {\bf 54,} R14257 (1996);
P.~Jaksch, I.~Yakimenko, and K.-F.~Berggren, Phys. Rev. B {\bf 74,}
235320 (2006); A.~A.~Starikov, I.~I.~Yakimenko, and K.-F.~Berggren,
Phys. Rev. B {\bf 67,} 235319 (2003).

\bibitem{BruusPE2001} H.~Bruus, V.~V.~Cheianov, and K.~Flensberg, Physica E (Amsterdam) {\bf 10,} 97 (2001).

\bibitem{MeirPRL2002} Y.~Meir, K.~Hirose, and N.~S.~Wingreen,
Phys. Rev. Lett. {\bf 89,} 196802 (2002); K.~Hirose, Y.~Meir, and
N.~S.~Wingreen, Phys. Rev. Lett. {\bf 90,} 026804 (2003).

\bibitem{MatveevPRL2004} K.~A.~Matveev, Phys. Rev. Lett. {\bf 92,} 106801 (2004).

\bibitem{ReillyPRB2005} D.~J.~Reilly, Phys. Rev. B {\bf 72,} 033309 (2005).

%\bibitem{JakschPRB2006} P.~Jaksch, I.~Yakimenko, and K.-F.~Berggren,
%Phys. Rev. B {\bf 74,} 235320 (2006);  A.~A.~Starikov, I.~I.~Yakimenko,
%and K.-F.~Berggren, Phys. Rev. B {\bf 67,} 235319 (2003).

\bibitem{RejecNature2006} T.~Rejec and Y.~Meir, Nature {\bf 44,} 2900 (2006). 

\bibitem{GolubPRL2006} A.~Golub, T.~Aono, and Y.~Meir, Phys. Rev. Lett. {\bf 97,} 186801 (2006).

\bibitem{LasslPRB2007} A.~Lassl, P.~Schlagheck, and K.~Richter, Phys. Rev. B {\bf 75,} 045346 (2007). 

\bibitem{SzaferPRL1989} Although a similar structure is known to occur
depending on the constriction shape of the QPC due to the resonance
[A. Szafer and A.~D.Stone, Phys. Rev. Lett. {\bf 62,} 300 (1989)], this
mechanism is not enough to explain the present anomaly problem.

\bibitem{BlanterPR2000} Y.~M.~Blanter, M.~B{\"u}ttiker, Phys. Rep. {\bf 336,} 1 (2000).

\bibitem{ReznikovPRL1995} M.~Reznikov, M.~Heiblum, H.~Shrtikman, and
D.~Mahalu, Phys. Rev. Lett. {\bf 75,} 3340 (1995). 

\bibitem{KumarPRL1996} A.~Kumar {\it et al.,} Phys. Rev. Lett. {\bf 76,} 2778 (1996).

\bibitem{LiuNature1998} R.~C.~Liu, B.~Odom, Y.~Yamamoto, and S.~Tarucha, Nature {\bf 391,} 263 (1998). 

\bibitem{GershonPRL2008} G.~Gershon {\it et al.,} Phys. Rev. Lett. {\bf 101,} 016803 (2008).

\bibitem{HashisakaPRB2008} M.~Hashisaka {\it et al.,} Phys. Rev. B {\bf
78,} 241303(R) (2008); Physica Status Solidi (c) {\bf 5,} 182 (2008).

\bibitem{LesovikJETP1989} G. B. Lesovik, Pis'ma
	Zh. Eksp. Teor. Fiz. {\bf 49,} 513 (1989) [JETP Lett. {\bf 49,}
	592 (1989)].

\bibitem{ButtikerPRL1990} M. B{\"u}ttiker, Phys. Rev. Lett. {\bf 65,} 2901
	(1990); M. B{\"u}ttiker, Phys. Rev. B {\bf 46,} 12485 (1992).

\bibitem{MartinPRB1992} Th. Martin and R. Landauer, Phys. Rev. B {\bf 45,} 1742 (1992).

\bibitem{ButtikerPRB1990} M.~B{\"u}ttiker, Phys. Rev. B {\bf 41,} 7906 (1990).

\bibitem{CommentOnHeating} The deviation of the observed Fano factor
from the theory for $G>G_0$ is attributed to the electron
heating~\cite{KumarPRL1996}.

\bibitem{SaddlePointNote} As the potential shape defined by the gate
electrodes is smooth on 2DEG the harmonic potential approximation is expected to be valid
in the present QPC to give a correct energy scale of $\Delta\epsilon$.

\bibitem{RokhinsonPRL2006} L.~P.~Rokhinson, L.~N.~Pfeiffer, and
K.~W.~West, Phys. Rev. Lett. {\bf 96,} 156602 (2006); J. Phys.:
Condens. Matter {\bf 20,} 164212 (2008).

\end{thebibliography}
\end{document}